\pdfoutput=1
\documentclass[iop]{emulateapj}
\usepackage{hyperref}

\usepackage[usenames,dvipsnames]{color}
\usepackage{rotating}
\usepackage{epstopdf}
\usepackage{color}
\usepackage{multirow}
\usepackage[para,online,flushleft]{threeparttable}

\newcommand{\lsim}{\raise0.3ex\hbox{$<$}\kern-0.75em{\lower0.65ex\hbox{$\sim$}}}

\newcommand{\kms}{km\ s$^{-1}$}

\usepackage{fancyheadings}
\usepackage{amsmath}
\usepackage{amssymb}
\usepackage{subfigure}
\usepackage{graphics}
\usepackage{natbib}
\bibpunct{(}{)}{;}{a}{}{,}
\usepackage{appendix}

\usepackage{booktabs}

\def\apj    {{ApJ }}
\def\apjl   {{ApJL }}
\def\apjs   {{ApJS }}
\def\aj     {{AJ }}

\def\araa   {{ARA\&A }}

\def\mnras  {{MNRAS }}
\def\nat    {{Nature }}

\def\pasj   {{PASJ }}

\begin{document}
\title{Trojans in the solar neighborhood}

\shorttitle{The Hercules stream}
\shortauthors{D'Onghia \& Aguerri}

\author{Elena D'Onghia\altaffilmark{1,}\altaffilmark{2} \& 
J.~Alfonso L.~Aguerri\altaffilmark{3,}\altaffilmark{4}}

\altaffiltext{1}{Department of Astronomy, University of Wisconsin, 475 North Charter
  Street, Madison, WI 53706, USA (edonghia@astro.wisc.edu)}
  
\altaffiltext{2}{Center for Computational Astrophysics, Flatiron Institute, 162 Fifth Avenue, New York, NY 10010, USA}

\altaffiltext{3}{Instituto de Astrofisica de Canarias, Calle Via Lactea, 38205, La Laguna, Spain}

\altaffiltext{4}{Departamento de Astrofisica, Universidad de La Laguna, Avenida Astrofisico Francisco Sanchez s/n, 38206 La Laguna, Spain}


\begin{abstract}
About 20\% of stars in the solar vicinity are in the Hercules stream, a bundle of stars that move 
together with a velocity distinct from the Sun. Its origin is still uncertain. Here, we explore the possibility that Hercules is made of {\it trojans}, stars captured at L4, one the Lagrangian points of the stellar bar. Using GALAKOS--a high-resolution N-body simulation of the Galactic disk--we follow the motions of stars 
in the co-rotating frame of the bar and confirm previous studies on Hercules being formed by stars in co-rotation resonance with the bar. Unlike previous work, we demonstrate that the retrograde nature 
of trojan orbits causes the asymmetry in the radial velocity distribution, typical of Hercules in the solar vicinity. We show that trojans remain at capture for only a finite amount of time, before escaping L4 without being captured again. We anticipate that in the kinematic plane the Hercules stream will de-populate along the bar major axis and be visible at azimuthal angles behind the solar vicinity with a peak towards L4. This test can exclude the OLR origin of the Hercules stream and be validated by Gaia DR3 and DR4.

\end{abstract}


\section{Introduction}
\label{sec:intro}
The Gaia satellite is mapping the phase-space of the Milky Way, revealing proper motions and 
distances for a few millions of stars in the solar neighborhood. The Gaia DR2 observations confirmed that stars 
in the solar vicinity do not have a smooth distribution of velocities: about 3.5 million stars within 
1.5 kpc of the Sun have a rich kinematic substructure that manifest as ridges in the velocity 
space \citep{Katz2018, Ramos2018}.

In particular, the distribution 
of stars in 
the velocity space defined by their radial and azimuthal velocities is dominated by a number of 
streams, bundles of stars that move together in the same direction with velocities that are distinct 
from neighboring stars \citep{Antoja2018}. 

In addition to the traditional phase-space analysis used in previous works to identify moving groups of stars 
in the solar vicinity, more recent investigations include the action-space analysis. The phase-space of star motions is six-dimensional and commonly described by stars' positions and velocities. The radial action, $J_{R}$ quantifies the amount of oscillation that the star exhibits in a 
radial direction along its orbit. In particular, this quantity describes the orbit's eccentricity. 
The azimuthal action $J_{\phi}=L_z$ is the angular momentum on the vertical component $z$. 
In axisymmetric potentials, all of these actions are integrals of motions; that is, they are constant 
along the orbits and therefore can be used to label different stellar orbits \citep{BT2008}. In the presence of strong 
perturbations, such as those caused by spiral arms and bars, the actions are not fully 
conserved (see e.g. \citealt{VeraCiroDonghia2016}). 

At a given point in time, however, the actions can still be calculated and may represent a diagnostic of 
non-axisymmetric perturbations \citep{1973Bible}. 
This is because long-lived perturbations like spiral arms and the 
stellar bar can lead to an orbital diffusion that forms distinct features in orbit space. Therefore, 
in the solar vicinity, streams and structures in the velocity space manifest as structures and ridges 
in the action-space \citep{Trick2018,Monari2019}. 

The origin of the structure in the phace-space is uncertain. Some works in the past pointed toward and extragalactic origin. Thus, 
\citet{Helmi1999} suggested that substructure in the kinematic plane 
was caused by accreted and disrupted satellites. However, the dynamical origin is the most accepted theory for 
the substructure in the kinematic plane.
As previously noted, these features cannot simply have arisen 
from groups of stars that were born with similar kinematics, because these streams have various ages and 
metallicities, as confirmed in detailed studies \citep{Famaey2005, Bensby2017}, and most likely 
have a dynamical origin, e.g. caused
by non-axisymmetric features of the stellar disk like the stellar bar and spiral arms.

While the Milky Way is a barred Galaxy \citep{deVacouleurs1964,Weinberg1992} with a spiral structure, there is 
still debate on the structure and morphology of the stellar bar, its orientation with respect to the Sun, and its 
rotation rate, usually referred to as its pattern speed. Also uncertain is the number and strength of 
the spiral arms that propagate in the Galaxy \citep{2014Reid}. The existence of perturbations in the Milky Way 
in the 
form of the stellar bar and spiral arms ensures that stars may have their orbits perturbed when they 
move at the same frequency as these features, a process called resonance. For a stellar bar, there are 
three strong resonances: i) the co-rotation resonance characterized by star's orbit moving at the same 
azimuthal frequency as the bar; ii) the inner-; and iii) the outer-Lindblad resonances, characterized 
by two radial oscillations for every azimuthal oscillation in the frame rotating with the Galactic bar 
pattern. The same resonances can be defined for spiral patterns.

\begin{figure*}[th] 
   \centering
   \includegraphics[width=7in]{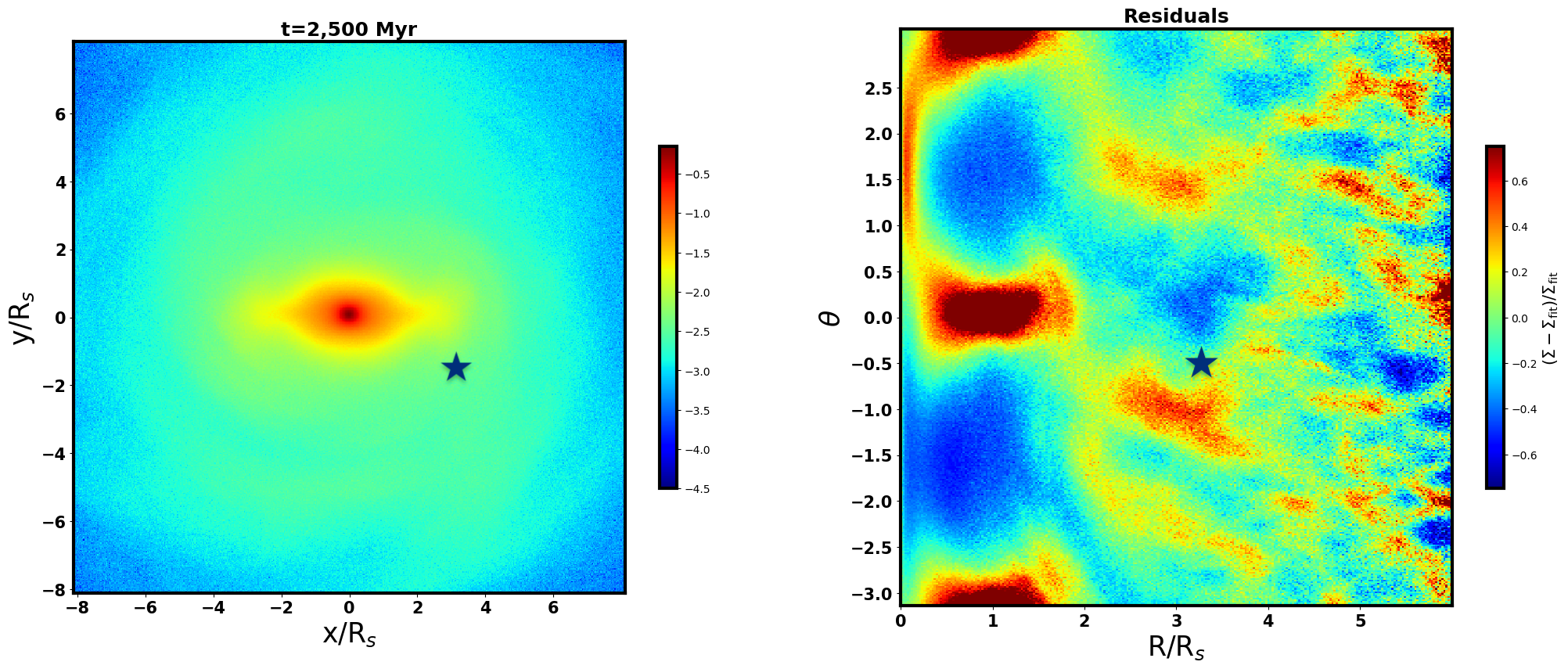} 
   \caption{The stellar disk displayed face-on in Cartesian coordinates after 2.5 Gyrs of evolution. A region of 30 kpc on a side is displayed (left panel) in units of the disk scale length $R_s$=2.6 kpc. The Sun is located at 8.1 kpc from the Galactic center at the azimuthal angle of -28$^{o}$ with respect the semi-major axis of the stellar bar (marked with a star). Surface density residuals are shown in polar coordinates (right panel). The bar extends as a horizonthal ridge in the central part of the disk. Two arms extend at the edge of the bar.}
\label{fig:FaceOnDisk}
\end{figure*}

Previous models of the inner Galaxy suggest that the radius of the outer-Lindblad resonance (OLR) of 
the Galactic bar lies in the vicinity of the Sun and affects the velocity distribution of nearby 
stars \citep{Kalnajs1991,Dehnen2000}.
These studies have shown that for a model resembling the stellar disk with a bar with pattern speed 
$\Omega_P=55$ km s$^{-1}$ kpc$^{-1}$, the OLR causes a bimodal 
distribution of the velocity between a moving group of low-velocity stars centered on the Local 
Standard of Rest (LSR) and an association of stars, the Hercules stream, moving outward and rotating 
more slowly than the Sun. The scenario proposed for the formation of the Hercules stream of stars 
matches the data if the Galactic bar has a radius of 3 kpc and is a fast rotator \citep{Monari2017,Fragkoudi2019}. 

Recent 
measurements based on the three-dimensional density of red-clump giants, however, show that the Galactic bar extends to $\sim$5 kpc from the Galactic center \citep{Wegg2015}, in agreement with previous estimates \citep{Benjamin2005}  
with a pattern speed of $\Omega_P \sim 39$ km s$^{-1}$ kpc$^{-1}$ \citep{Portail2017}. These values were recently confirmed 
by using a modified version of the \citet{TW1984} method by combining proper motions of Gaia DR2 and the VVV
survey \citep{Sanders2019,Clarke2019}. If this 
is the case, then the OLR is placed at ~10.5 kpc from the Galactic center, too far for the Hercules 
stream of stars to be caused by orbits migrating outwards at this resonance \citep{PerezVillegas2017}. 
The former study showed that for a bar with pattern-speed of ~39 \kms kpc$^{-1}$ the Hercules stream 
consists of stars trapped at co-rotation resonance with the bar.

Other studies reproduce distributions of stars with a similar degree of substructure in the kinematic space by invoking a sequence of 
short-lived spiral transients \citep{DeSimone2004,QuillenMinchev2005,Quillen2018}, but the Hercules stream is not always 
well matched.

In an attempt to better reproduce the observed positions and velocities (namely the phase-space) of 
the stars in the Hercules stream, more recent models have included both the stellar bar and the spiral 
structure of the Milky Way. These studies, however, in some cases are based on N-body simulations that 
reproduce the solar vicinity by stacking several snapshots owed to the 
lack the spatial resolution of a few hundred parsecs \citep{Fragkoudi2019}. Other studies adopted semi-analytical 
models in which the spiral 
structure and the bar are modeled as externally applied perturbations with assumed 
properties \citep{Hunt2018Arms,Monari2019}. 

Here, following \cite{PerezVillegas2017}, we approach the question of the formation and evolution of 
the Hercules stream in the long-bar scenario. We set up a high-resolution N-body simulation of the
 Milky Way with structural parameters that reproduce the currently observed properties of our 
Galaxy \citep{Ortwin} and with enough spatial resolution to resolve the solar vicinity. 
This includes a self-consistent spiral pattern and bar whose properties are in agreement with current  
observations.
The advantage of this methodology is that the spiral waves and the bar arise from the 
stars themselves as self-consistent collective disturbances \citep{Donghia2013,Donghia2015}. 
In fact, if self-consistency is not an 
integral part of the model, the amplitude of the perturbation and the time-dependence assumptions made 
in the model may bias the resonant interactions between the bar, the spiral arms, and the stars, which 
are responsible for the formation of the Hercules stream.

The structure of this paper is as follows: we describe our methodology in Section 2. In Section 3 we present our results by showing the evolution of the Hercules stream by time. An example of the orbit of a star trapped in  co-rotation with the bar and its escape 
are also illustrated. Finally, validating tests of our model are provided.

\begin{figure}[th] 
   \centering
   \includegraphics[width=3in]{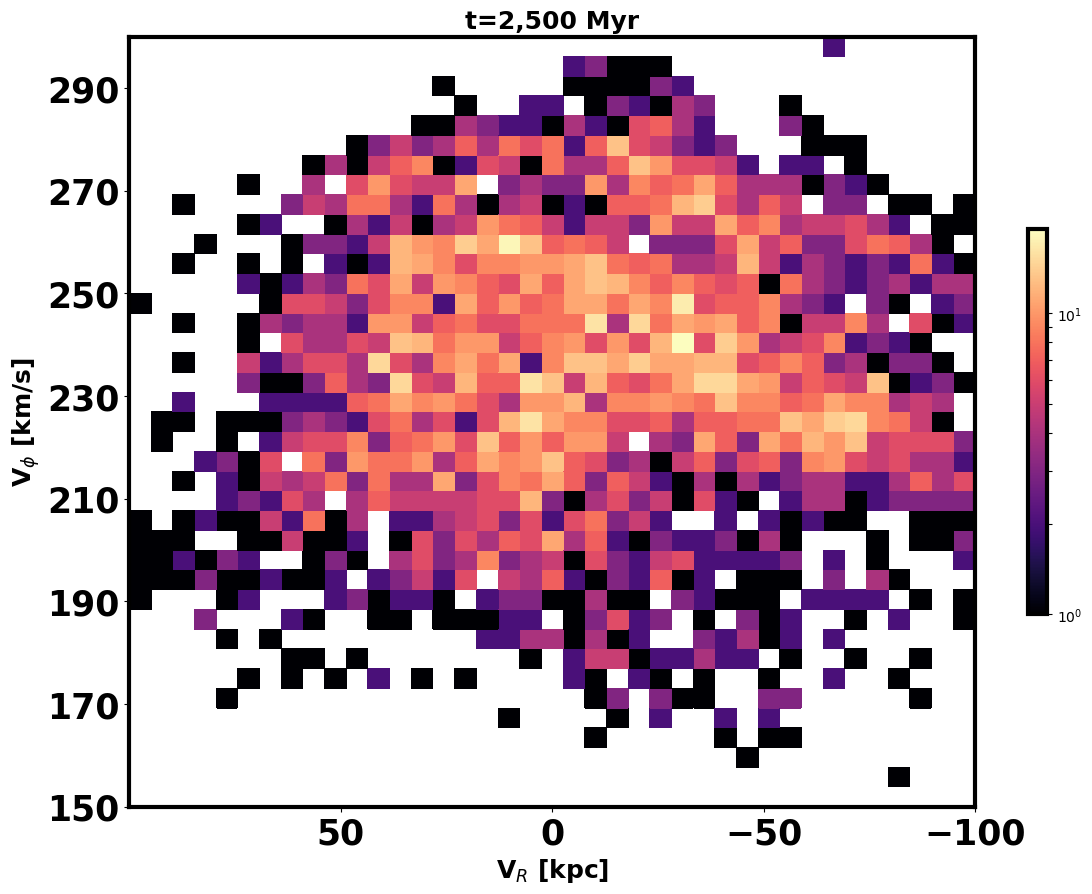} 
   \caption{A bi-dimensional histogram of the number of stars in the solar vicinity is displayed in the velocity
plane. Note that, in agreement with the Gaia DR2 data, the velocity plane is inhomogeneous, revealing signatures of 
substructures in kinematic space. }
\end{figure}

\section{Methods}
\label{sec:methods}
The simulations were carried out with GADGET3, a parallel TreePM-Smoothed particle hydrodynamics (SPH) code 
developed to compute the evolution of stars and dark matter, which are treated as collisionless fluids. The 
six-dimensional phase space is discretized into fluid elements that are computationally realized as 
particles in the simulations. A detailed description of the code is available in the 
literature. Here we note its essential features. 

GADGET3 is a cosmological code in which the gravitational field on large scales is calculated with a 
particle-mesh algorithm, while the short-range forces are computed using a tree-based hierarchical 
multipole  expansion, resulting in an accurate and fast gravitational solver. This scheme combines the high 
spatial resolution and relative insensitivity to clustering of tree algorithms with the speed and accuracy 
of the particle-mesh method to calculate the long-range gravitational field. 

Pairwise particle interactions are softened with a spline kernel of scale-length $h_{s}$, so that they are 
strictly Newtonian for particles separated by more than $h_{s}$. The resulting force is 
roughly similar to traditional Plummer softening with scale-length $\epsilon=h_{s}/2.8$.
For our applications the gravitational softening length is fixed to 40 pc for the dark halo and 28 pc for 
the stellar disk and 80 pc for the bulge. The total number of N-Body particles employed in the simulation 
(stellar disk, bulge, and halo) is approximately 90 million (see the Appendix for the details).

\subsection{Solar vicinity and velocity definitions}
In this simulation, named GALAKOS, the Sun is located at 8.1 kpc from the Galactic center at the azimuthal angle of -28$^{o}$ 
with respect the semi-major axis of the bar (see Figure 1). In cartesian coordinates its in-plane location is identified at (x,y)=(7.15, -3.8) kpc. Within 300 pc at the Sun's location there are more than 4,000
stars  with -200 $\le$ z $\le$ 200 pc. The stellar bar rotates counter-clockwise, and thus the Sun 
moves behind it. After 2.5 Gyrs of the life of the Milky Way, the stellar bar is in place with a length of approximately 4.5 kpc and a pattern speed of 40 \kms kpc$^{-1}$ (see the Appendix for details), 
in full agreement with current data \citep{Ortwin}. Note that in the disk there are two strong arms extending
from the edge of the bar to 5 scale lengths in radius ($\sim 11$ kpc). As shown in Figure 1 the stellar disk
shows deviations from being axisymmetric with an asymmetry in the amplitude of the two spiral arms dominating
the disk. 

\subsection{Action-angle space}
In an axisymmetric disk, the energy $E$ and the angular momentum $L_z$ are the classical integrals for 
the in-plane motion. Actions are an alternative set of integrals of motion. For motion in 
the symmetry plane of an axisymmetric potential $\Phi(R,z)$ the equation of the action function $S$ 
separates in two terms, one azimuthal and one radial respectively: $S = S_{\phi}(\phi) + S_{R}(R)$
\citep{1973Bible}.

Then 
\begin{equation}
S_R=\int^R \dot{R} dR
\end{equation}
and 
\begin{equation}
    S_{\phi}=L_z \phi
\end{equation}
The radial action is:
\begin{equation}
J_R=\frac{1}{2 \pi} \int \frac{\partial{S}}{\partial{R}} dR
\end{equation}
\noindent
where the integral is taken over the full period.
The azimuthal action is:
\begin{equation}
J_{\phi}=\frac{1}{2 \pi} \int \frac{\partial{S}}{\partial{\phi}} d\phi= L_z
\end{equation}
\noindent
$J_{\phi}=L_z$ is the specific angular momentum of the star around the Galactic center.
Angles, $\theta_{\phi}$ and $\theta_{R}$ can be conveniently defined and are the conjugate variables to the actions.
Angles increase at uniform rate: $\dot{\theta}_{\phi}=\Omega_{\phi}$, the azimuthal frequency and 
$\dot{\theta}_{R}=\Omega_{R}$, the radial frequency.

When the angle and action variables are used $J_{\phi}$ may be considered as the radial area of the 
orbit, $J_{R}$ as the extent of the oscillation, $\theta_{\phi}$ as pointing at the angular position of the
epicenter around the Galaxy and $\theta_R$ as pointing at the phase of the epicyclic oscillation
about that epicenter.

\begin{figure*}[th] 
   \centering
   \includegraphics[width=7in]{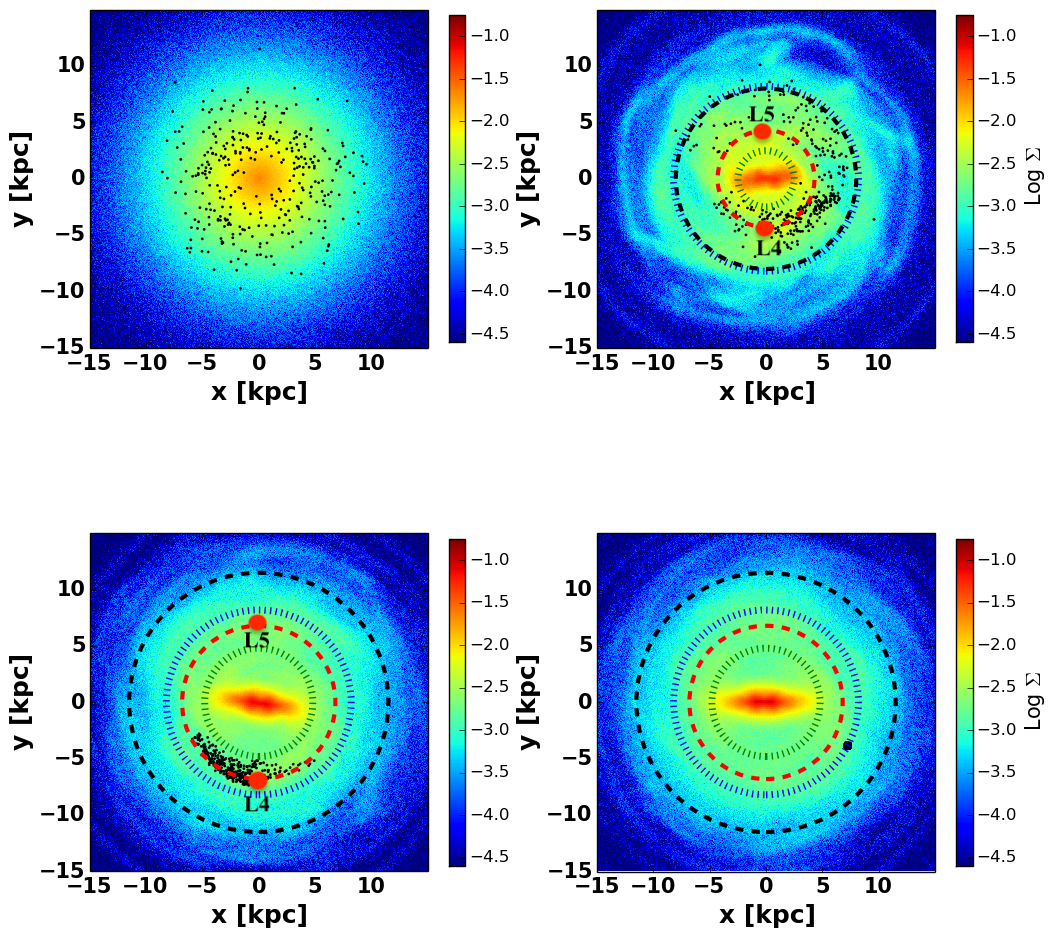} 
   \caption{The evolution of the Hercules stream over time. Current Hercules stream stars are selected in
the neighborhood of the Sun with V$_{\phi}$ between 180 and 220 \kms and $-50 \le V_r \le 100$ \kms according to 
Figure 2 and the Gaia DR2. 
Their positions are traced back in time and displayed in the co-rotating frame of the bar. Stars are
initially distributed following the exponential distribution of disk stars (left top panel). After
1 Gyrs, the Hercules stars still cross 
through the edge of the bar (right top panel). After 2 Gyrs, stars are in co-rotation with the bar
librating around the Lagrangian point L4 (bottom left panel). The bar rotates in the counter-clockwise direction. At each time, the radius where bar resonances occur is displayed: the co-rotation radius (red dashed line),
together with L4 and L5, the inner ultra-harmonics (green circle), the OLR (black circle), and the solar
radius (blue circle). The Hercules stars at the current time are marked by the black patch (right bottom
panel). For visualization, only a subsample of stars is displayed.}
\end{figure*}

We used the AGAMA software package \citep{Vasiliev2019} to extract the 
potential from each snapshot of our simulation and 
compute the actions and angles (see Appendix for details). 


\section{Results}
\subsection{The velocity plane}

An inspection of the radial and azimuthal velocity of nearby stars in Galactocentric coordinates 
shows that  the solar neighborhood is far from smooth in its velocity distribution. 
Figure 2 displays an underlying smooth component of stars but the entire stellar 
distribution shows evidence of substructures in kinematics, which are streams of stars
with coherent velocities. 
Note that the Hercules stream is reproduced in our model as a structure extending with azimuthal velocity 
V$_{\phi}$=180-220 \kms and radial velocity $-50 \le V_r \le 100$ \kms. 
According to this definition 18\% of stars at 
the solar vicinity in GALAKOS populate the Hercules stream. Thus, as compared to the Sun's motion, Hercules is moving slower and shows an asymmetry pronounced in the radial velocity distribution with more stars  moving away from the Galactic center as compared to the ones moving inwards as indicated also by the data \citep{Dehnen2000, Antoja2018}.
In what follows, we investigate whether this kinematic feature in the velocity space is the result of dynamical processes, such as the effects of the bar and spiral arms, which are fully captured in our simulation.

\subsection{Donkey orbits}

In the frame of a rotating bar, the effective potential is given by:

\begin{equation}
    \Phi_{eff}=\Phi-\frac{1}{2} \Omega_P^2R^2
\end{equation}

\begin{figure*}[th]
 \centering
 \includegraphics[width=7in]{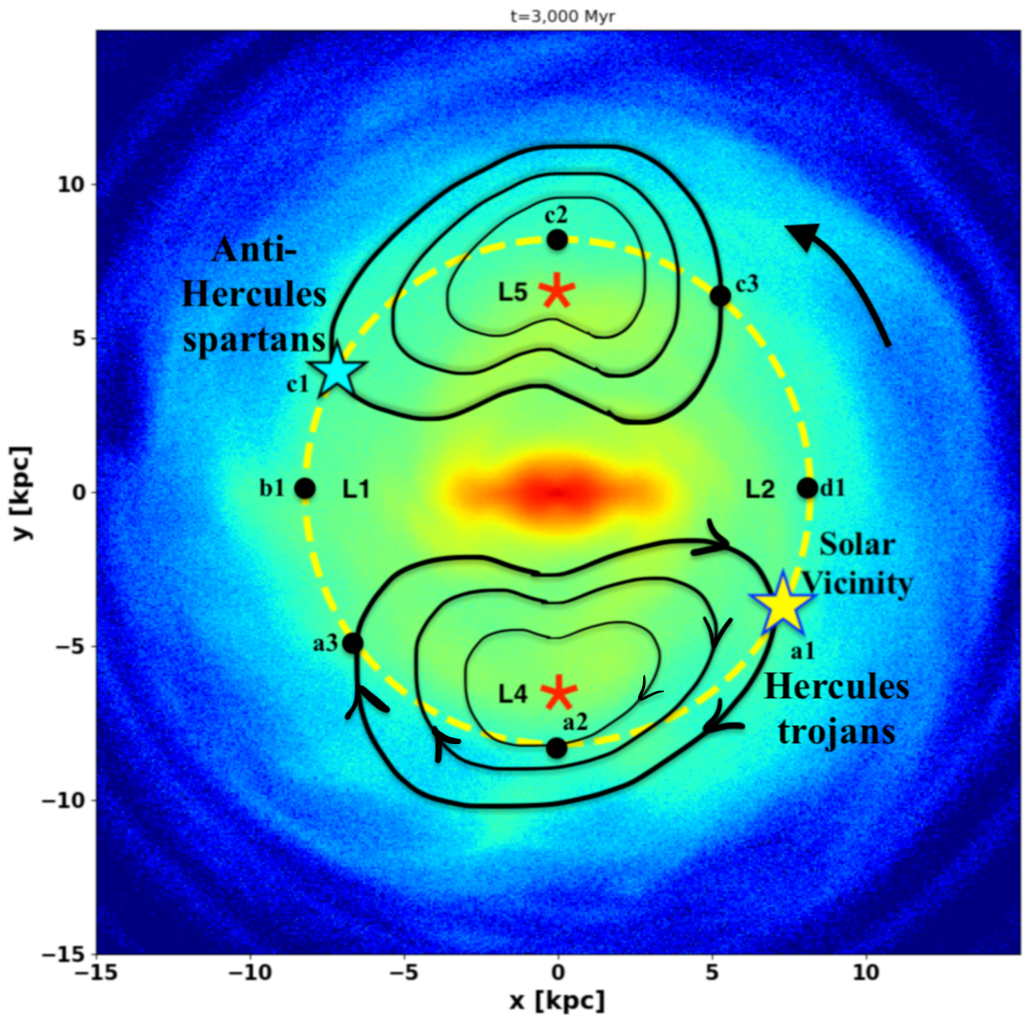}
 \caption{A schematic view of the stellar disk.
Face-on view of the stellar disk in the co-rotating frame of 
the stellar bar (at the center) rotating counter-clockwise. The Hercules stream is in the Solar vicinity
(marked with the yellow star), with -28$^{o}$ of displacement with respect the major axis of the bar. 
The azimuthal angles along the solar radius are: $\phi=0$ (labeled with $d1$), $\phi=-28^{o}$ ($a1$), 
$\phi=-90^{o}$ ($a2$), $\phi=-152^{o}$ ($a3$). 
The Hercules stars are librating around the Lagrangian point L4. A stream with the same kinematic features as
Hercules will be located at an azimuth 180$^{o}$ from the Sun's vicinity (marked with the blue star). The
anti-Hercules consists of spartans trapped in co-rotation with the bar
librating around the opposite Lagrangian point L5. The azimuthal angles of relevance for anti-Hercules are:
$\phi=180^{o}$ (labeled with $b1$), $\phi=28^{o}$ ($c3$), 
$\phi=90^{o}$ ($c2$), $\phi=152^{o}$ ($c1$). 
Curves of constant effective potential are drawn
schematically with solid black lines as bounding curves with a given energy. As the energy increases the
bounding curves grow larger. The solar radius is displayed at 8.1 kpc from the Galactic center (drawn with
the yellow dashed line).
 }
 \label{cartoon}
\end{figure*}
with $\Phi$ the total potential and $\Omega_P$ the pattern speed of the bar. In barred galaxies, the Lagrangian
points--defined as points in the plane where the effective potential is minimal or maximal--play an important role.
In general, bars have two unstable Lagrangian points, L1 and L2, at the minimum potential shortly beyond their ends, along the
major axis, and two other points, L4 and L5, along the intermediate axis at the maximum of potential 
(labeled in Figures 3-4). The situation is similar to that encountered in the restricted three-body problem, except that there is no equilibrium point L3. Such a characteristic means that rotation is an essential factor in bar dynamics \citep{deVacouleurs1964,Pfenniger1990}.

The disk of the Galaxy, as described by an axisymmetric exponential radial profile, is dominated by orbits
that appear in the x-y plane to be rosettes of varying eccentricities, and with the morphology dictated
by the radial energy. Such regular orbits have constant phase-space coordinates and are considered
integrable. 
These orbits can be represented by a combination of three fundamental frequencies: $\Omega_{\phi}$, 
$\Omega_r$, $\Omega_z$ the azimuthal, the radial, and the vertical frequency, respectively. 
The Jeans equations have been
used to perform simplistic assessments of the orbital structure of real galaxies under the assumption that
galaxies can be described by the classical integrals of motion, dictated by $E$ and angular momentum $L_z$.
Unfortunately this condition does not apply to dynamical studies of realistic galaxies. Indeed, 
commensurate (or resonant) orbits arise when a perturbation, represented by a stellar bar or spiral arms, is 
applied \citep{Petersen2019}.

For nearly circular orbits the radial frequency is the epicycle frequency, $\Omega_r=\kappa (R)$, while 
$\Omega_{\phi}=\Omega (R)$, the circular 
motion. Resonances occur when the frequency at which a star encounters the perturbation potential
resonates with one of the natural frequencies $\kappa$, $\Omega$ of the stellar orbit.
When the stellar orbits are displayed in the frame rotating with the angular velocity of the perturbation, represented by the arms
or the bar, then the perturbation will affect only the orbits which close in this frame. When this condition occurs, 
the bar and arms perturb the star for long time to change the stellar orbit significantly \citep{1973Bible}. Resonances occur when:

\begin{equation}
m (\Omega_{\phi}-\Omega_P)=\pm l \Omega_r
\end{equation}

where $l$ is a positive, negative or zero integer and $m=2$ in presence of the bar. In the co-rotation resonance $l=0$ stars move in phase with the bar $\Omega_{\phi}=\Omega_P$. 
First we compute the fundamental frequencies of the stars identified in the velocity plane as Hercules in Figure 2 
and verified that their $(\Omega_{\phi}-\Omega_P)$/$\Omega_P <<1$. After 2.5 Gyrs of evolution all stars in the Hercules stream satisfy the condition to be {\it near} co-rotation with the stellar bar. 

Once we identified the Hercules stream, we traced its position back in time and followed its stars' 
orbits in the frame co-rotating with the bar (Figure 3). Stars that at the final time are in the solar vicinity 
(marked with black points in the right lower panel) were initially distributed following the
exponential distribution of disk stars (left top panel). 

Note that at 1 Gyrs, the bar is already in place and 
the majority of Hercules stars {\it circulate} the disk, crossing through the edge
of the bar (right top panel). 

Those stars feel the torques from the pattern potential in their orbits around the Galaxy. Since they are not aligned with the bar these torques 
add up over several orbital times. Only after long time those stellar orbits will respond to  
the perturbation and become {\it trojans}. Then they are trapped in resonant orbits in co-rotation with the bar but librate around the equilibrium point, L4, which is located 90$^{o}$ away from the bar.  
Although those orbits feel the torques from the bar pattern potential, they are incapable of aligning with it. 
Donald Lynden-Bell called  them {\it donkey} because of their inability to cooperate with a potential well of the perturbation and deepen it by aligning with it. These stars lose and gain angular momentum depending on their location with respect the bar.  When close to the bar and moving ahead of the bar, stars are being pulled backwards by the bar and lose angular momentum \citep{Ceverino2007}. At the same time, stars that are behind the bar are being pushed forwards gaining
angular momentum. However those orbits are never cooperative and will never align with the bar potential 
\citep{Earn1996}. 

The bar torque with a two-fold symmetry is the mechanism that induces the Hercules stars to librate around L4. 
Therefore, Hercules stars have two-fold symmetric orbits with symmetry of 180$^o$ in an axisymmetric potential of a stellar bar. This implies that by symmetry there is an anti-Hercules stream consisting of stars librating around L5, that we name  {\it spartans} by analogy to moon and minor planets in the solar system. 
Figure 4 illustrates the situation with the Sun that will encounter in 100 Myrs anti-Hercules, the twin stream located 180$^{o}$ from the solar vicinity.

Note that trojans around L4 cannot librate around L5. This is a difference with previous studies claiming that the Hercules stream consists of stars librating either around L4 or L5 or circulating 
with the bar \citep{PerezVillegas2017}. However, the former study does not account for a time-varying potential, a limitation that might have affected their stellar orbits. 

\begin{figure}[h] 
   \centering
   \includegraphics[width=3in]{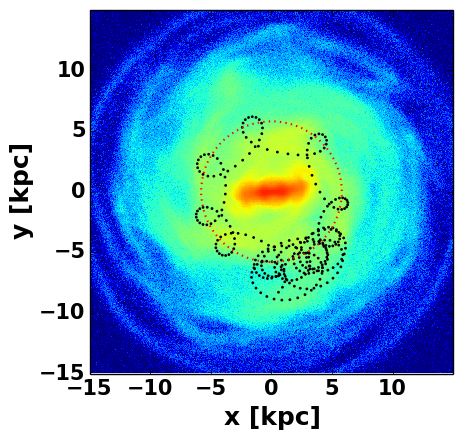}
   \caption{Evolution in time of the orbit of an Hercules star from 1,3 to 2.5 Gyrs of evolution. 
   The orbit is displayed in the rest frame of the bar.
   A star on a retrograde orbit, {\it donkey orbit}, with a period of 10$^8$ yrs circulates the disk.
   The dotted red circle indicate the co-rotation radius of the bar.}
\end{figure}

\begin{figure}[h] 
   \centering
   \includegraphics[width=3in]{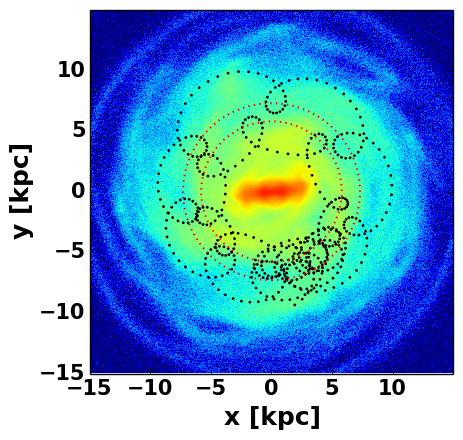}
   \caption{The full evolution in time of the orbit of a Hercules star. The star is first trapped in co-rotation with the bar and librates around L4. In the next 500 Myrs of evolution the star escapes L4 migrating to a larger radius and continues its 
   circulation around the Galactic center. The dotted red circles indicate the co-rotation radius of the bar at earlier (inner circle) and later (outer circle) time.}
\end{figure}

\begin{figure*}[th] 
   \centering
   \includegraphics[width=6in]{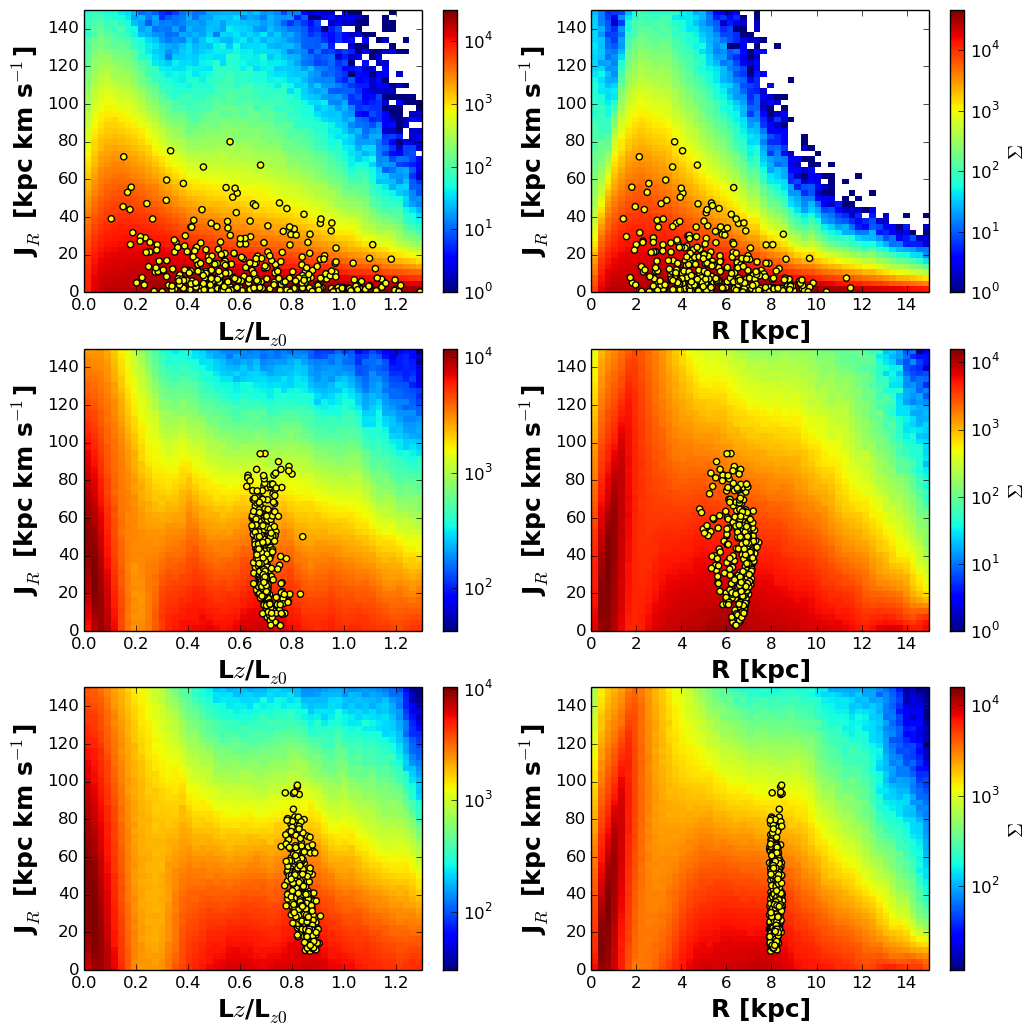} 
   \caption{Time evolution of the action space of the stellar disk. {\it Left Side}.
The bi-dimensional histogram of the radial and azimuthal actions in the stellar disk computed within
15 kpc on a side are displayed at the initial time of the disk evolution (t=0, top panel), after 2 Gyrs 
(middle panels) and at the final time (2.5 Gyrs) (bottom panel). The Hercules stars are displayed (marked with yellow filled circles). The angular momentum
of the stars is normalized to the angular momentum of the Sun today. {\it Right Side}. Time evolution of the
radial action of the stellar disk displayed against the cylindrical radius. Symbols are the same as in
the left panels.}
\end{figure*}

We cannot display the orbits of all Hercules stars in this paper. We shall therefore provide an example 
of the orbit of a Hercules star in Figures 5-6 followed for 4 Gyrs of evolution in the frame co-rotating with the bar. The star is moving on a retrograde orbit with a period of approximately 
10$^8$ yrs as illustrated in Figure 5. After 2 Gyrs the star on a periodic orbit gets 
trapped in a librational motion around L4, with a period of 250 Myrs. Thus, stars in co-rotation resonance 
with the bar are on retrograde orbits. Their motion can be described by (see Pichon, the Knight prize 1992, \citealt{Pichon1993}): 

\begin{equation}
    \vec{v}=- \vec{\nabla} \Phi_{eff} + 2 \vec{\Omega} \ \wedge \ \vec{v}
\end{equation}

Note that the retrograde periodic orbit is stable even if the orbit is far from L4. 
However, as the energy increases the amplitude of the orbit around 
the equilibrium points grows until the star escapes from L4 and continues in its circulation around the center of the Galaxy. As an example to illustrate this point,
Figure 6 shows the case of the same star on a periodic orbit in the frame co-rotating with the bar followed for 
4 Gyrs of evolution. It becomes a trojan after 2 Gyrs of the disk evolution and at current time librates around L4. In the next 500 Myrs of evolution the star escapes from L4 and moves to a more energetic orbit
continuing its circulation across the stellar disk (displayed as the more external retrograde orbit in Figure 6). 

As the co-rotation moves outward
passing the solar circle we expect the Hercules stream to still be visible at the solar vicinity but
its stars will be on orbits with larger guiding radii, hence greater angular momentum than the Sun. Thus, in the kinematic plane they are expected to populate the region with greater V$_{\phi}$ and 
will appear as they move inward the Galactic center ($V_r \le 0$) in Figure 2.

The amount of trapped matter near L4 that reaches the solar vicinity after 2.5 Gyrs of evolution 
is ~18\%. Other numerical experiments suggest an estimate of ~4\% (M. Weinberg, private communications) for a barred galaxy. It is unclear at this stage whether differences in the initial conditions lead to different estimate of the trapped matter
in numerical simulations. 

\begin{figure*}[th!] 
   \centering
   \includegraphics[width=7in]{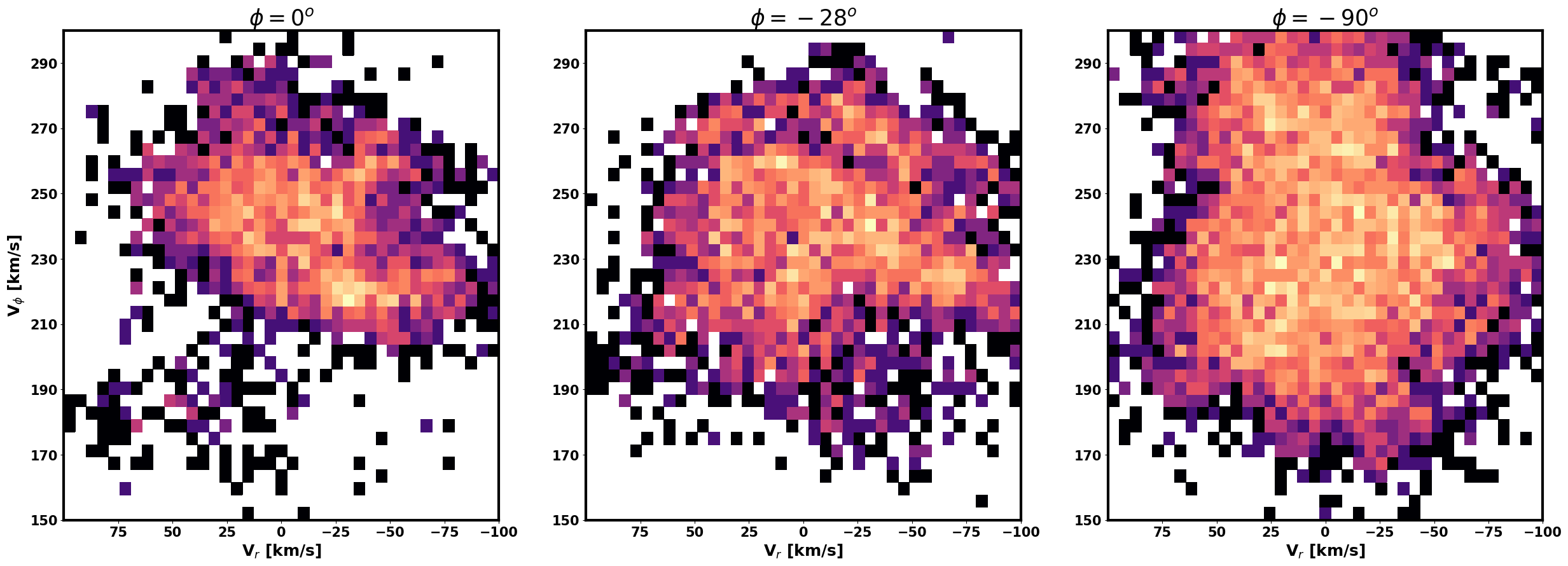} 
   \includegraphics[width=7in]{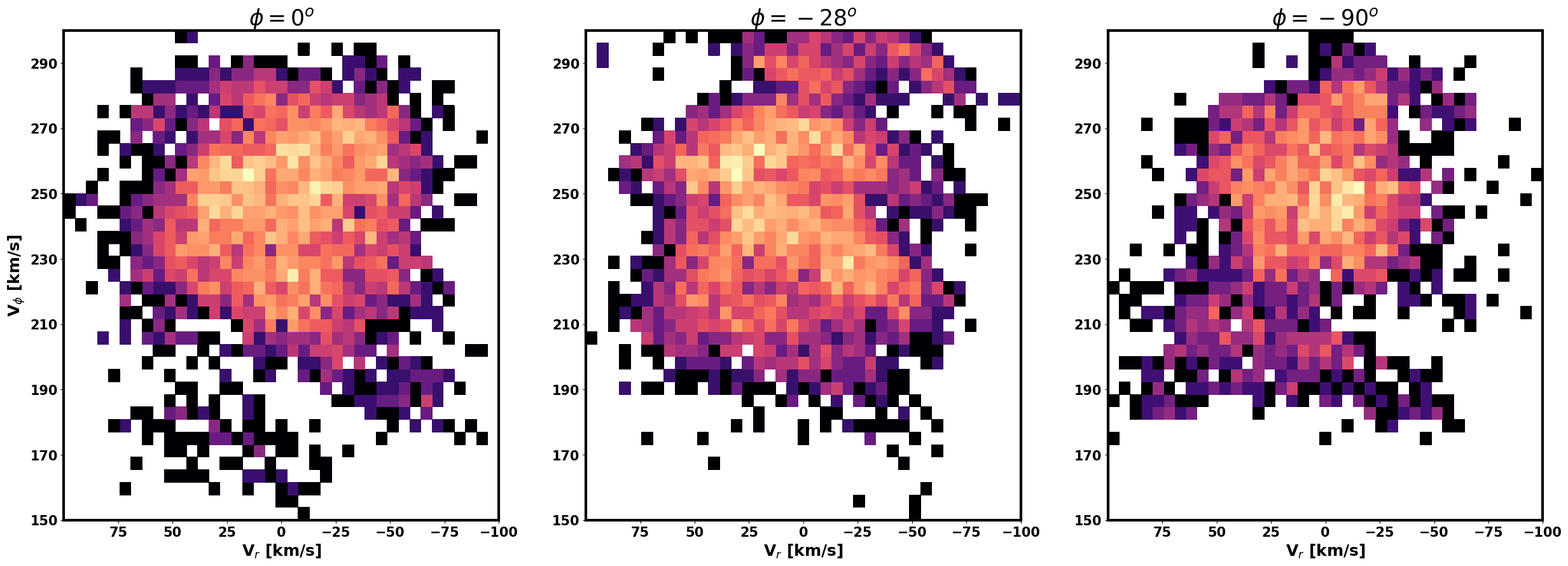} 
   \caption{{\it Top Panels.} The V$_{\phi}$-V$_r$ plane for stars selected in patches of 300 pc each at various azimuthal angles around the solar radius (R=8.1kpc). Trojans disappear for patches ahead of the Sun along the bar semi-major axis of the bar in proximity of L2 (left panel). Along the intermediate axis, at $\phi=-90^{o}$ degrees, more Hercules stars are observed with lower azimuthal velocity (right panel) as compared to the solar vicinity (middle panel). {\it Bottom Panels.} The velocity distribution for stars selected in patches of 300 pc each at the same  azimuthal angles as above panels but  around R=8.6 kpc after 1.5 Gyrs of evolution of the disk. The OLR is at 7.8 kpc at this time (see Table 2 in the Appendix) so that R$_{OLR}/$R$_{\odot}$=0.9 as the fiducial case of \citet{Dehnen2000}, Figure 2.} 
\end{figure*}

\subsection{Escape from L4}
A question concerns the eccentricities of the orbits of Hercules stars. The observational data suggest that Hercules 
stars have eccentric orbits \citep{Dehnen2000,Trick2018}. However
according to the perturbative theory applied to stellar orbits, 
when the motion of stars is perturbed the angles can be split in slow and fast 
variables 
$\theta_{slow}=\theta_{\phi}-\Omega_P t$ is the angle from the Galactic center
to the orbit's epicenter as measured in axes rotating with the bar.
The period of an orbit is small compared with its precession rate relative to the rotating perturbation.
This assumption allows averaging over the fast motion of the star around its orbit. 
Formally, averaging over the fast motion of the stars produces an exactly constant {\it fast action}: 
\citep{LyndenBellKalnajs1972,1973Bible,LyndenBell1979}:

\begin{equation}
    J_{fast}=J_R - \frac{l}{m} J_{\phi}.
\end{equation}

Thus, in co-rotation (l=0), $J_{fast}=J_R$ is the adiabatic invariant of the motion around that epicenter.
Why do trojans have eccentric orbits?

First, we define Hercules stars in the kinematic plane and those are {\it near} co-rotation with the bar. Second, we notice that some stars that will be trojans are born already with eccentric orbits (see Figure 7 top panels). 
Before being trapped in co-rotation resonance with the bar some stars are increasing the random energy of their orbits by the passage of four spiral arms
that in our simulation form before the bar. The increase of Hercules eccentricity follows the in-plane heating of the disk that occurs when the non-axisymmetric features develop. In fact the sequence of panels of Figure 7 (top and middle) shows that in the inner part of the disk there is a sharp increase of the radial action of stars that does  not significantly change their guiding radius, indicating in-plane disk heating. This occurs when the spiral structure propagates and subsequently the bar starts forming.  

\begin{figure*}[th!] 
  \centering
   \includegraphics[width=3.5in]{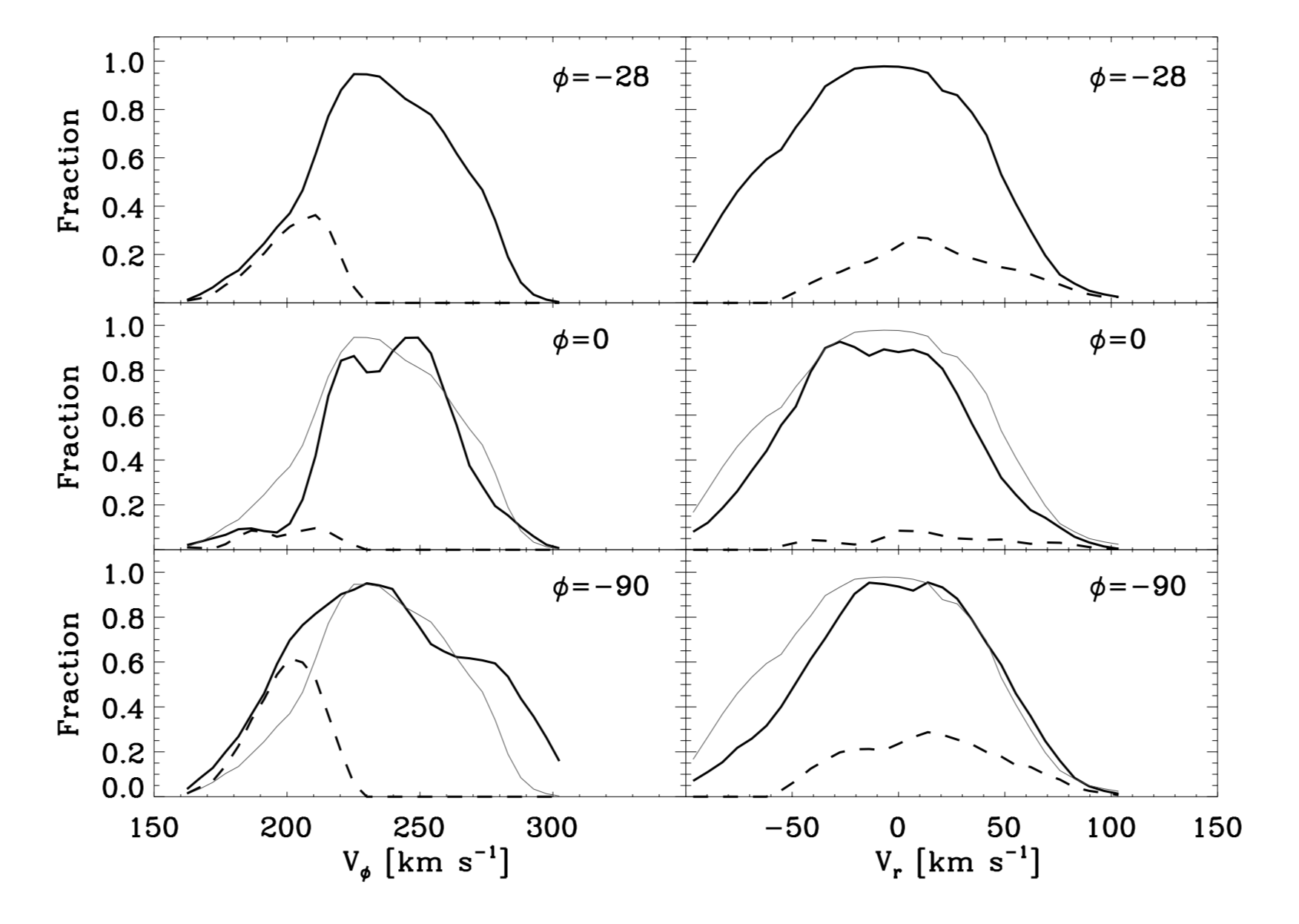}
   \includegraphics[width=3.5in]{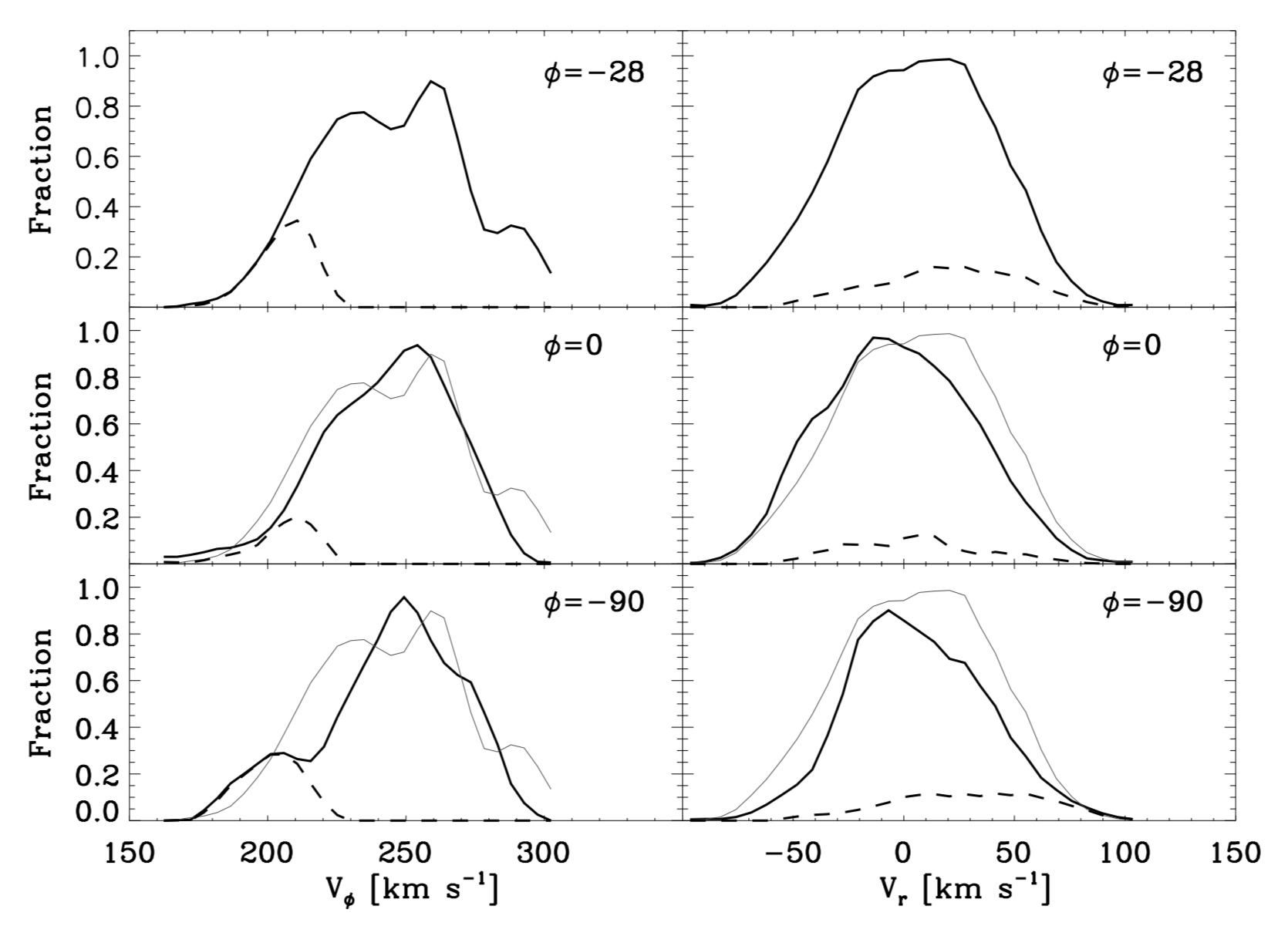}
   \caption{{\it Left Side}. Azimuthal and radial velocity distributions are 
displayed for the Hercules stream (dashed lines) at various azimuthal angles (from top to down) at the solar radius (8.1 kpc) for a long-bar scenario.
The total velocity distribution of all stars in the patch at each azimuthal angle is displayed (solid black line). The total velocity distribution at the solar vicinity (the case of $-28^{o}$) is overlapped at each panel (gray solid line). {\it Right Side}. The azimuthal and radial velocity distribution are shown for the case of a short-fast bar
with the OLR located at 7.8 kpc.}
\end{figure*}

Once trojans complete their exchange of angular momentum with the bar, they can produce a very large 
libration around L4 that allows stars to reach the solar vicinity (see also \citealt{PerezVillegas2017}).
As they are on very energetic orbits around L4 they escape from the Lagrangian point.
As L4 moves out to larger radii, trojans that 
are trapped move out to larger guiding radii. In fact, as the bar slows in its pattern speed, the co-rotation 
region moves outwards and the stars that were trapped in co-rotation migrate radially outwards too.

To illustrate this situation, Figure 7 displays the radial action, $J_r$, of the Hercules stream 
(yellow circles) as a function of their guiding radii (represented by $L_z$) as a function of time and as 
compared to the radial and azimuthal actions of the stars of the disk. At the initial time, the Hercules 
stars have the same actions as the stars in the background (top left panel). After 2 Gyrs of the 
evolution of the stellar disk, more and more Hercules stars have guiding radii locked at the co-rotation radius 
with the stellar bar, located approximately at 5.5 kpc (middle left panel). After another 500 Myrs (at the current time) 
of evolution the pattern speed of the bar lowered by 5 \kms kpc$^{-1}$ and the co-rotation radius moved out to a galactic radius of 6.0 kpc 
and Hercules stars that were trapped with the guiding radii at the co-rotation of the bar migrate to larger radii
(see Table 2 in the Appendix for the evolution of Resonances locations over time).

Note that the right panels in Figure 7 show the time evolution of the radial 
action of the stellar disk, displayed against the cylindrical radius with the Hercules stream overlapped 
and consisting of stars that at the current time have cylindric radii in the solar vicinity (at 8.1 kpc 
from the Galactic center).

\subsection{Limitation of the Model}
This numerical experiment suffers from some limitations that we plan to improve in future. First, while the dark halo, 
the stellar disk and bulge are live, no gas component is included in the Galactic disk. The lack of gas likely affects the 
evolution of the bar as gas should take part in the angular momentum exchange between the dark halo and the disk. 
\citet{Athanassoula2013} demonstrated how the gas fraction affects the bar properties. Galaxies with large 
fractions tend to have shorter and weaker bars. 
Second, the bulge is described with a Hernquist model and does not rotate. This limitation accelerates 
the exchange of angular momentum between the bar and the other components of the Galaxy leading to a fast lowering of the bar pattern speed and consequent rapid movement of the resonances outward.  
The description of the evolution of the trojans hold nevertheless, although the details of the times at capture at L4 and escape will depend on the Galaxy potential adopted.

\begin{table}[h]
\caption{
Fraction of Hercules stars measured at various patches along the solar circle for 
a long and a short-fast bar.}
\centering
\begin{tabular}{| c | c | c |}
\hline 
  $\phi$       &  \% Hercules       &  \% Hercules \\
               &   Long bar          &  Short bar (R$_{OLR}/$R$_{\odot}=0.9$)\\
\hline
  0$^{o}$      &   6\%              &      10\% \\ 
\hline
  -28$^{o}$    &   18\%             &      14\%  \\   
\hline
  -90$^{o}$    &   25\%             &      16\%  \\
 \hline
\end{tabular}
\end{table}

\subsection{Validating tests}

The experimantal evidence in support of this model concerns the kinematic plane, in particular the estimate of the V$_{\phi}$-V$_r$ at the solar circle at various azimuthal angles that will be possible to validate with Gaia DR3 and DR4. Figure 8 (top row) illustrates the predictions for V$_{\phi}-$V$_r$ of stars selected in patches of 300 pc each at $\phi=0;-28^{o};-90^{o}$ in the long-bar model. Trojans  will de-populate in patches ahead of the Sun along the bar semi-major axis in proximity of the Lagrangian point L2 (left top panel for $\phi=0$). 
Because trojans will mostly populate the regions around  L4, for a patch of stars selected along the bar semi-minor axis (at $\phi=-90^{o}$), more Hercules stars will have low V$_{\phi}$ in the $V_{\phi}$-V$_r$ 
plane (right top panel of Figure 8) as compared to the values in the solar vicinity (middle top panel). Note that this behavior 
seems not to be found in models of a short and fast-rotating bar with the OLR located near the solar radius as displayed in Figure 8 (bottom panels). In this situation Hercules consists of stars changing their orbits from x$_{1}(1)$ to x$_{1}(2)$  \citep{Dehnen2000}. Hercules does not increase significantly  
at $\phi=-90^{o}$. In this case the velocity distributions are evaluated after 1.5 Gyrs of evolution 
of the disk, when the bar pattern speed is 55 \kms kpc$^{-1}$, and the bar length is 3 kpc. 
The OLR is located at 7.8 kpc (see Table 2 in the Appendix). The Sun is placed at 8.6 kpc in order to compare 
to the fiducial case of Figure 2 of \citet{Dehnen2000} with R$_{OLR}/R_{\odot}$=0.9.

A quantitative estimate is displayed in Figure 9. Azimuthal, V$_{\phi}$, and radial velocity, V$_{r}$, 
distributions are shown for the Hercules stream in the long-bar scenario (left side, dashed lines) and short-bar scenario (right side, dashed lines). The fraction of stars in Hercules, 
defined as V$_{\phi}$=180-220 \kms and -50 $\le$ V$_{r} \le$100 \kms,  is 18\% in the solar vicinity, 24\% towards the bar semi-minor axis and decreases to 6\% for a patch of stars along the bar semi-major axis (see Table 1). 
While in both scenarios the fraction of stars in Hercules is similar in the solar vicinity and at patches selected along the bar major axis, Hercules is expected to be more prominent along the bar
semi-minor axis in the case where the bar is long as compared to the case where it is short and fast rotating.

In the solar vicinity Hercules shows an asymmetry in the $V_r$ distribution in both scenarios as 
displayed in Figure 9. However, in the case of the long bar the asymmetry can be interpreted by having  
an excess of trojans with V$_r \ge$0. In fact those stars are on donkey orbits in the co-rotating frame of the bar and librate from the center of the Galaxy on a retrograde motion with respect the Sun. Therefore more Hercules stars come from the inner part of the Galaxy as compared to the ones that move towards the Galactic center. The radial velocity distribution in Figure 9 (left side) breaks
in the proximity of L4 ($\phi=-90^{o}$) because at that location the same amount of 
trojans move inwards or outwards. 

For a perfectly axisymmetric disk the asymmetry in the $V_{\phi}$-V$_r$ plane will be 
inverted in the quadrant for $\phi=-90^{o}$ to $\phi=-180^{o}$. Symmetric predictions hold for anti-Hercules formed by stars librating around L5. Finally, Hercules is expected to be less visible at larger Galactic radii, as fewer and fewer trojans will be captured at L4 on such energetic orbits.

\section{conclusions}
In this paper, we examined the properties of the Hercules stream at the solar vicinity in the long-bar scenario.  
We carried out GALAKOS, a high-resolution N-body simulation with enough spatial resolution to resolve 
the solar neighborhood. The simulation consists of the Milky Way model that accounts for a time-varying potential that after 2.5 Gyrs of evolution forms a bar with a length of 4.5 kpc and a 
pattern speed of 40 \kms kpc$^{-1}$.  While our analysis has much in common with earlier studies of the Hercules stream, it differs in the following respects.

\begin{enumerate}
    \item [1.] In our N-body simulation the Galactic potential is time-varying. Density waves that characterize the spiral 
    structure and the bar arise from the stars themselves as self-consistent collective disturbances with amplitude that changes by time. We confirm previous studies that Hercules stars are on trapped orbits in co-rotation with the bar \citep{PerezVillegas2017,Monari2019}. However, our results indicate that Hercules is entirely made of trojans, stars that librate only around L4, the stable Lagrangian point located 90 degrees perpendicular to the major axis of the bar. Stars trapped around L5, the other stable Lagrangian point, that also end up as part of Hercules in \citet{PerezVillegas2017}, belong in our model 
    to anti-Hercules, which is expected to be located  at 180$^{o}$ from the Sun. Those stars are not visible 
    in the solar vicinity. 

\item[2.] Trojan evolution is dictated by the evolution of the bar and the Galactic potential. 
   In our model they last at capture at L4 only 500 Myrs on very energetic orbits, before escaping without being captured again. 
   
\item[3.] Our theory explains the asymmetry of the radial velocity distribution of Hercules. Because these stars are on donkey orbits, they are retrograde in the frame co-rotating with the bar. At the Sun's location of -28$^{o}$ of inclination with respect the bar, Hercules will appear as a stream with the majority of stars coming from the inner part of the Galaxy, with positive radial velocity V$_{r}$. The asymmetry is expected to break for patches
of stars selected towards the bar semi-minor axis (L4).

\item[4.] Finally, we suggest a test in the kinematic plane that allows us to distinguish between Hercules being 
stars at the outer-Lindblad resonance (OLR) of the bar or trojans in co-rotation with the bar. 
Trojans  will de-populate in patches of stars along the solar circle but align with the major axis of the bar in the proximity of L2 ($\phi=0$). At -90$^{o}$ towards L4 the fraction of Hercules will be 24\% of stars of the patch 
and will have lower azimuthal velocity as comparted to the solar neighborhood. This increased population 
of Hercules stars towards the semi-minor axis is not expected in models of a short and fast-rotating bar with the OLR located near the solar radius. This anticipation is a valid test for the forthcoming Gaia DR3 and DR4.

\end{enumerate}

\acknowledgments{
E.D.O is grateful to A. Kalnajs and K. Freeman for much generous advise during her visit at the ANU
as Stromlo Distinguished Visitor of the Research School of Astronomy and Astrophysics (RSAA). 
The authors are grateful to C. Pichon, M. Weinberg and J. Navarro for many insightful suggestions. E.D.O. acknowledges support from the Sierra Foundation at the Instituto de Astrofisica de Canarias (IAC) in Tenerife,  the Center for Computational Astrophysics at Flatiron Institute,
the Institute for Theory and Computation (ITC) at Harvard, and the KITP program for the hospitality during the completition of this work. This research was supported in part at KITP by the National Science Foundation under Grant No. NSF PHY-1748958, and the grant from the Spanish Ministerio de Economía y Competitividad (MINECO) AYA2017-83204-P. GALAKOS is run at the La Palma supercomputer center. 

}

\newpage
\appendix

\subsection{Setting the Initial Conditions for GALAKOS}
This section explains our mass models and describes the realization of the initial conditions. We begin 
with the details of our fiducial simulation. The Milky Way in our study (GALAKOS) consists of a dark matter 
halo, a rotationally supported disk of stars, and a spherical stellar bulge. The parameters describing each 
component are independent and models are constructed in a manner similar to the approach described in 
previous work \cite{SpringelII2005}.

\begin{figure*}[th!] 
   \centering
   \includegraphics[width=3.5in]{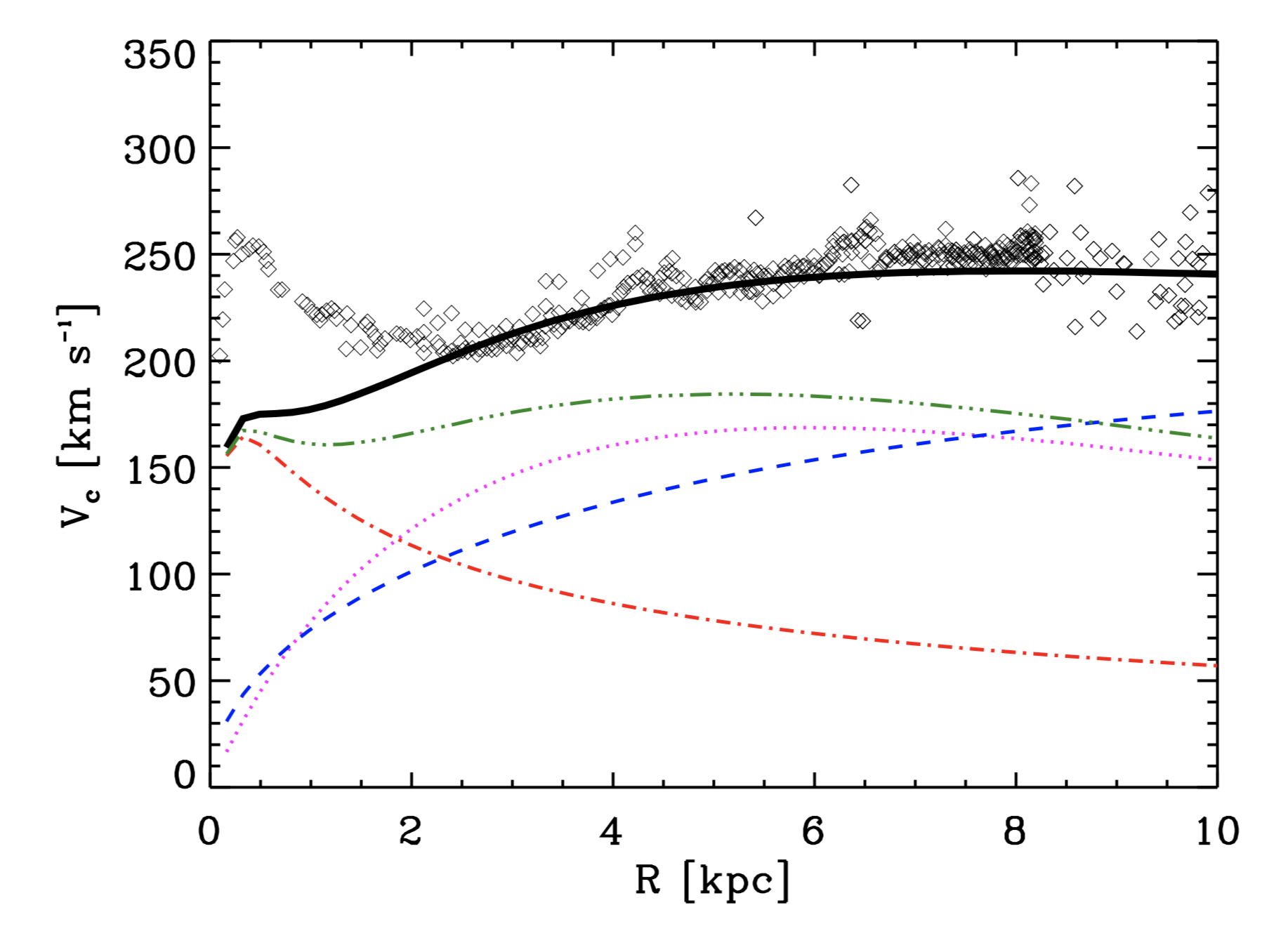}
    \includegraphics[width=3.5in]{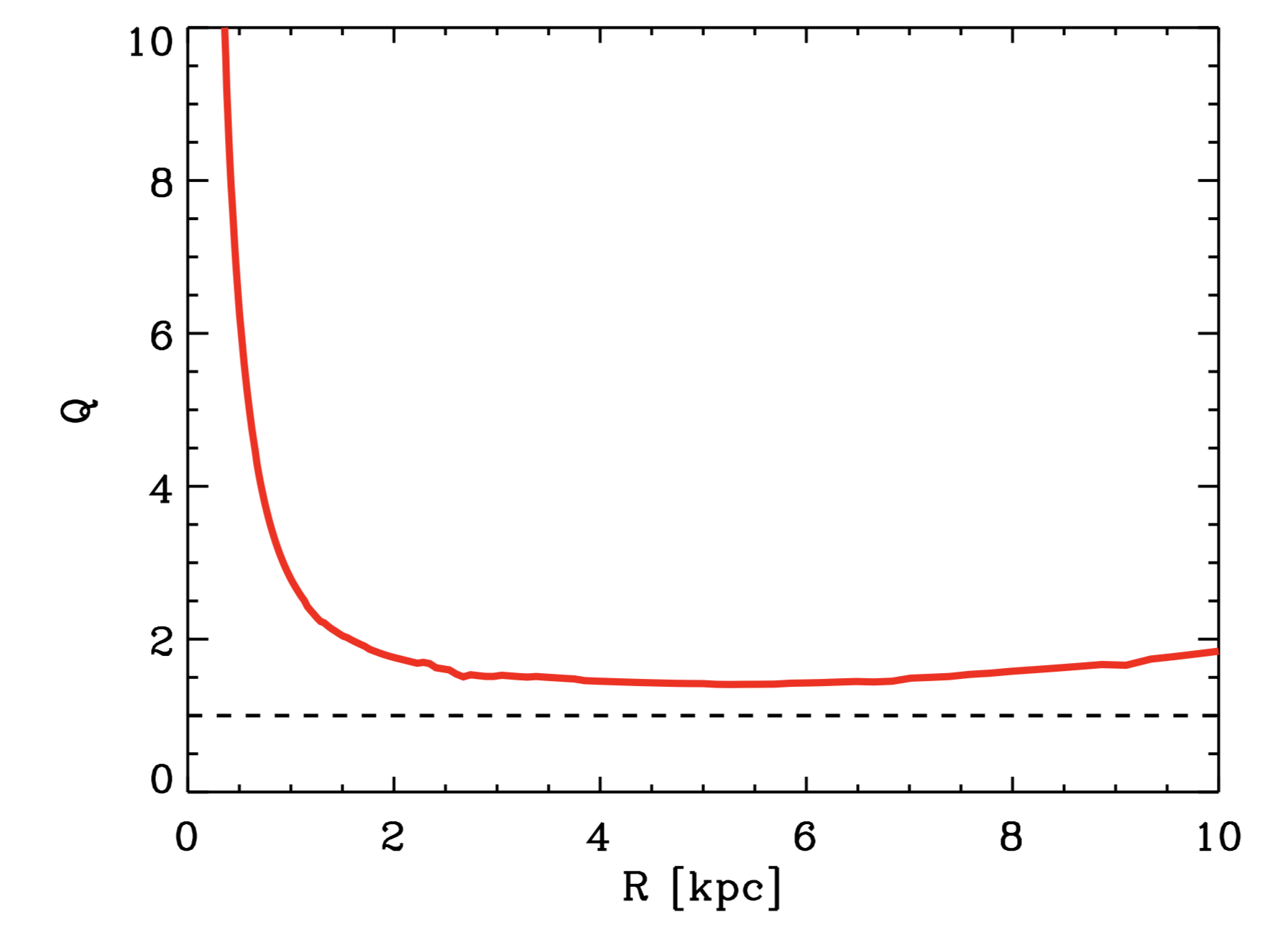}
   \caption{Structural parameters of GALAKOS. {\it Left Side}. Initial velocity curves of GALAKOS.
The following components of the Galaxy are displayed: the dark halo (marked with a blue dashed line),
the stellar bulge (dot-dashed red line), the stellar exponential disk (violet dotted line). The stellar
rotation curve composed by the stellar disk and the bulge (dot-dashed green line) and the total rotation
curve with the stellar bulge, disk, and halo are also displayed (solid dark line). At the initial time the
bar is not present and therefore not accounted in the total rotation curve. The observed rotation curve of
the Milky Way \citep[][]{Sofue2009} is overlapped (marked with open black diamonds)
and matches the model between 4 and 8 kpc, the
range of radii where the data are more accurate. {\it Right Side}. The Toomre parameter for the stellar
disk at the initial time and during the evolution is larger than 1 across the stellar disk with a minimum
of 1.5.
 }
\end{figure*}

\subsection{Dark Matter Halo} 
We model the dark matter mass distribution of each galaxy with a Hernquist 
profile \citep{Hernquist1990}:  
\begin{equation}
\rho_{\rm{DM}}=\frac{\rm{M}_{\rm{DM}}}{2 \pi} \frac{a}{r(r+a)^3}
\end{equation}
\noindent
with cumulative distribution M($<r$)=M$_{\rm{DM}}r^2/(r+a)^2$ where $a$ is the radial scale length. 
This choice for the halo profile is motivated by the fact that in its inner parts, the shape of this 
profile is identical to the Navarro-Frenk-White fitting formula for the mass density distribution of 
dark matter halos in cosmological simulations. The total mass of the halo is
M$_{\rm{DM}}=$1x10$^{12}$ M$_{\odot}$ and it is sampled with 60 million particles. Moreover, the radial 
scale length for the halo is $a=30$ kpc. 

\subsection{Stellar Disk} For the simulation presented here, we adopted an exponential radial stellar disk 
profile with an isothermal vertical distribution:
\begin{equation}
\rho_d(R,z)=\frac{\rm{M}_d}{8 \pi z_0 \rm{R}_d^2} \rm{e}^{-\rm{R}/\rm{R}_d} \rm{sech}^2(\rm{z}/\rm{z}_0)
\end{equation} 
\noindent
where M$_d$ is the disk mass, R$_d$ is the disk scale length, and z$_0$ is the disk scale height. 
We set M$_d$=4.8x10$^{10}$M$_{\odot}$, R$_d$=2.67 kpc, and z$_0$=320 pc throughout in reasonable agreement 
with the values quoted in the literature for the observed structural parameters of the Milky 
Way \citep{Ortwin}. The disk  
mass is discretized with 24 million particles. The total baryonic mass (stellar disk and bulge) is 
M$_d$=5.6x10$^{10}$M$_{\odot}$. We do not include a gas component. The rotation curve has a disk 
fraction $f_D=V^2_c/V^2_{\rm{tot}}=0.54$ at R=2.2 R$_d$, the radius at which the exponential disk 
reaches the maximum circular velocity.

disk velocities are chosen by solving the Jeans equations in cylindrical coordinates in
the combined disk--halo potential. We set the radial velocity dispersion from the Toomre
stability equation:
\begin{equation}
\sigma_r(R)=Q \frac{3.36 G \Sigma(R)}{\kappa(R)}
\end{equation}  
\noindent

\begin{figure}[th!] 
   \centering
   \includegraphics[width=7in]{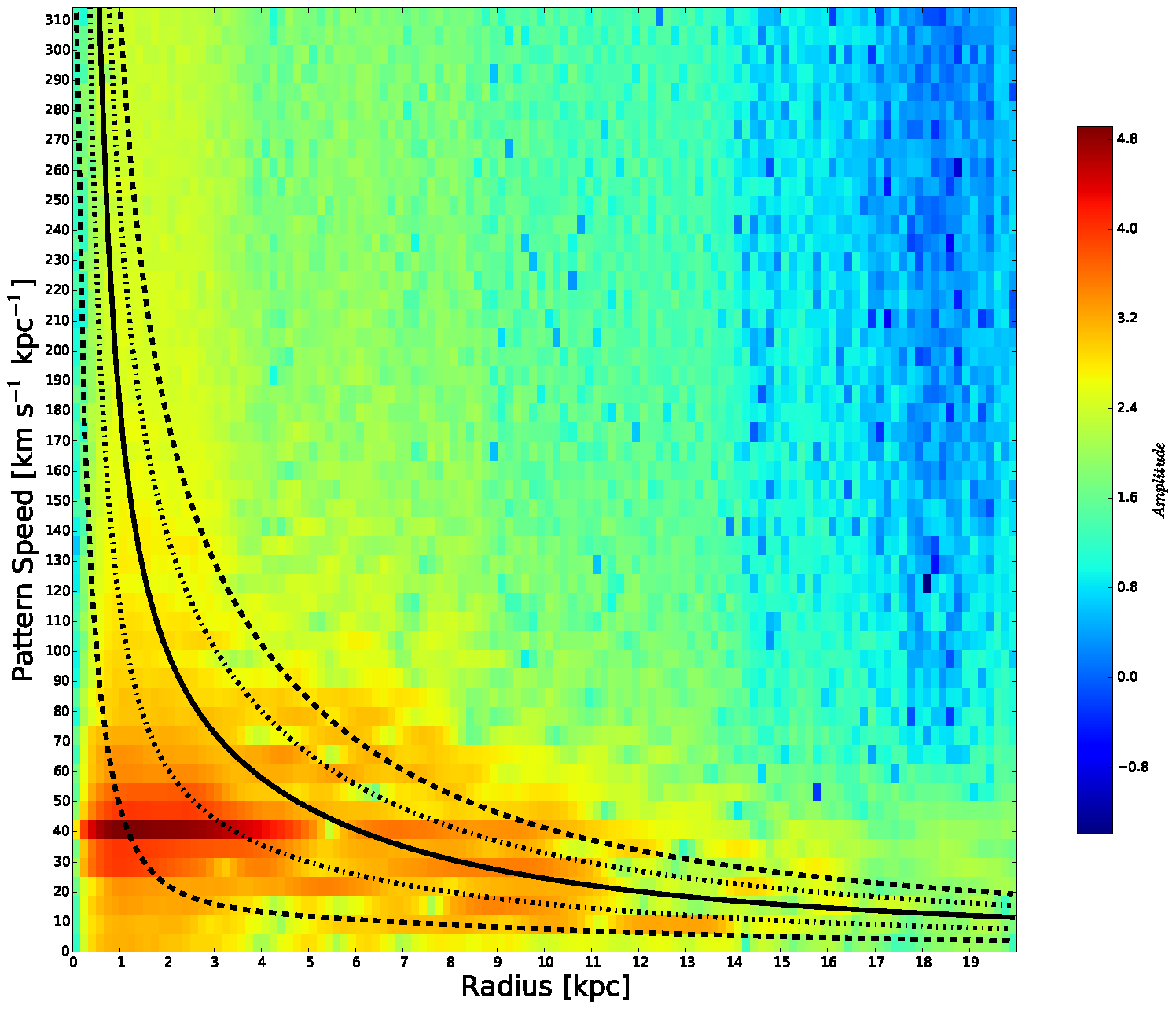}
   \caption{Power spectrum as a function of radius and frequencies for the $m=2$ 
Fourier Harmonic in the stellar disk. The strongest non-axisymmetric feature is the stellar bar with a
pattern speed of 40 \kms kpc$^{-1}$ at the current time. The solid line shows $m$ times the circular angular
frequency $\Omega=V_c/R$, which determines the co-rotation radius of the bar around 6 kpc, with $V_c$ 
the circular velocity. The long-dashed lines
correspond to Lindblad resonances. The dot-dashed lines display
the Outer-Ultra-harmonics resonances.}
   \label{fig:power_spectra}
\end{figure}

where G=1, $\Sigma(R)$ is the stellar surface density, and $\kappa$ the epicyclic frequency. The Q 
parameter reaches a minimum value of 1.5 between 3 and 7 kpc as displayed in Figure 10 
(right panel).

\subsection{Stellar Bulge}
The stellar bulge is described by the Hernquist model \citep{Hernquist1990}. The 
total mass of the bulge adopted in the 
simulation is M$_B$=8x10$^9$M$_{\odot}$ with a scale length $a_B$=120 pc.The number of particles sampled in 
the bulge is 8.4 million. The bulge in this numerical experiment does not rotate.

\subsection{Action and Angles Estimates}

We used the AGAMA software package \citep{Vasiliev2019} to extract the 
potential from each snapshot of our simulation and 
compute the actions and angles. We adopt the St{\"a}ckel fudge method \citep{SandersBinney2016} 
for computing actions and angles under the assumption that the motion is integrable and is locally well-described by a St{\"a}ckel potential, 
separable into prolate spheroidal coordinates. Actions and angles can be computed exactly for a general 
potential separable in a confocal ellipsoidal coordinate system. AGAMA includes the implementation of 
this approach. Here we note the essential procedure. A St{\"a}ckel potential assumes the form:
\begin{equation}
\Phi_S(u,v)=\frac{U(u)-V(v)}{\rm{sinh}^2u+sin^2v}
\end{equation}
\noindent
and is defined by two one-dimensional functions, U(u) and V(v), instead of being an arbitrary function of 
two coordinates. An orbit in such a potential respects three integrals of motion: energy (E), angular 
momentum (Lz), and a non-classical third integral. Given the three integrals, the equations defining the 
turn-around points (min/max values of u, v) can be numerically solved and then the actions, angles, and 
frequencies can be computed by one-dimensional numerical integration. The St{\"a}ckel fudge method uses the 
same expressions, but substitutes the real potential instead, which is not of a St{\"a}ckel form. Actions 
computed in this way are approximate, i.e., not conserved along a numerically integrated orbit. The 
accuracy of approximation depends on the only free parameter in the method: the focal distance. This 
parameter is obtained from the combination of second 
derivatives of the potential; hence this equation is implemented in the real potential at each input point.

\subsection{Spectral Analysis of the Stellar Disk}

The Galactic bar changes the internal structure of disk galaxies by rearranging the angular momenta and 
energy of star orbits that would otherwise be conserved. After 2.5 Gyrs of evolution, the bar is in place 
with a length of 4.5 kpc. Figure \ref{fig:power_spectra} displays the power spectra for $m=2$ harmonic mode, measured from snapshots 
sampled every 5 Myrs for the stellar disk in the 2,000$\le$ t/Myr $\le$ 2,500 interval. 
The solid line shows $m$ 
times the circular angular frequency $\Omega=V_c/R$, which determines the co-rotation radius of the bar 
located around 6 kpc. The ratio of co-rotation to the bar length in this 
model is 1.3. This value is similar to the estimates of other barred galaxies 
and indicates that the MW bar is fast-rotating \cite{Aguerri2015}.
At t=1.5 Gyrs the bar has a length of 3 kpc and a pattern speed of 55 \kms kpc$^{-1}$, while 
at 2.5 Gyrs its pattern speed is $\sim$40 \kms kpc$^{-1}$.

\begin{table}[h]
\centering
\caption{CR: co-rotation resonance, ILR: Inner Lindblad resonance,OLR: Outer Lindblad resonance,\\
IUH: Inner ultra-harmonic, OUH: outer ultra-harmonic.}  
\begin{tabular}{| c | c | c | c | c | c | c |}
\hline
 t [Gyrs] & $\Omega_P$ [\kms kpc$^{-1}$] & CR [kpc] & ILR [kpc] & OLR [kpc] & IUH [kpc] & OUH [kpc]\\ 
\hline
 1.5      & 55                  & 4.0       & 1.0      & 7.8      & 2.2   &  6.0 \\   
\hline
 2.5      & 40                  & 6.0       & 1.5      &  10.8    &  3.7  &  8.5 \\
 \hline
 3.0      & 35                  & 6.8       & 1.5       & 11.5      & 4.0  & 9.1 \\ 
\hline
 3.5      & 30                 &  7.5       & 1.5     & 12.0     & 5.0  &  10 \\
 \hline
\end{tabular}

\end{table}


\end{document}